# Sub-electron noise infrared camera development using Leonardo large format 2Kx2K SWIR LmAPD array

Philippe Feautrier [a], Jean-Luc Gach [a], Dan Owton [b], Matthew Hicks [b], Ian Baker [b], Keith Barnes [b], David Boutolleau [a]

[a] First Light Imaging SAS, Europarc Sainte Victoire, Bâtiment 5, Route de Valbrillant, Le Canet, 13590 Meyreuil, France

[b] Leonardo, First Avenue, Southampton, SO15 0LG, UK

## ABSTRACT

There have been no significant breakthroughs in infrared imagery since the hybridization of III-V or II-VI narrow-bandgap semiconductors on complementary metal-oxide semiconductor (CMOS) read-out integrated circuits (ROICs). The development of third-generation, linear-mode avalanche photodiode arrays (LmAPDs) using mercury cadmium telluride (MCT) has resulted in a significant sensitivity improvement for short-wave infrared (SWIR) imaging. The first dedicated LmAPD device, called SAPHIRA (320x256/24μm), was designed by Leonardo UK Ltd specifically for SWIR astronomical applications requiring speed and sensitivity. In the past decade there has been a significant development effort to make larger LmAPD arrays for low-background astronomy and advance adaptive optics.

Larger LmAPD formats for ultra-low noise/flux SWIR imaging, currently under development at Leonardo include a 512 x 512 LmAPD array funded by ESO, MPE and NRC Herzberg, a 1k x 1k array funded by NASA and a 2K x 2K device funded by ESA for general scientific imaging applications. The 2048x2048 pixel ROIC has a pitch of 15 microns, 4/8/16 outputs and a maximum frame rate of 10 Hz.

The ROIC characterization is scheduled in the third quarter of 2022, while the first arrays will be fabricated by end-2022. The hybridized arrays will be characterized during end-2022. At this time, First Light Imaging will start the development of an autonomous camera integrating this 2Kx2K LmAPD array, based on the unique experience from the C-RED One camera, the only commercial camera integrating the SAPHIRA SWIR LmAPD array. The main features of this camera is presented. The detector will be embedded in a compact high vacuum cryostat cooled with low vibration pulse at 50-80K which does not require external pumping. If necessary, an active vibration damping system can be added for reducing the array vibrations down to 0.01 micron. Sub-electron readout noise is expected to be achieved with high multiplication gain. Custom cold filters and beam aperture cold baffling will be integrated in the camera.

The large format 2Kx2K ROIC for APD array is funded by ESA under a TDE program with the contract number 000130154/20/NL/AR.

## 1. INTRODUCTION

Since 2008, Leonardo has been funded by the astronomy community (notably, the European Southern Observatory, ESO, and the Institute of Astronomy at the University of Hawai'i, UH) to develop new infrared detectors for the next generation of astronomical instruments. Initially the emphasis was to develop fast wavefront sensors for active programs, such as GRAVITY, however a number of breakthroughs in the APD technology has resulted in reduced dark current (by nine orders of magnitude) now making them candidates for astronomical imaging.

Conventional infrared detectors achieve a read noise of 10-15 rms electrons in a single double correlated sample reducing to 2.5 to 5 rms electrons by noise destructive readout and frame averaging. There is no imminent path to overcome the read noise limit in conventional detectors. HgCdTe avalanche photodiode arrays effectively reduce the read noise by noiseless avalanche multiplication of the signal charge within the photodiode. For astronomy APD gain delivers lower signal-to-noise and with enough avalanche gain the device becomes "noiseless". This type of APD is called linear-mode to separate it from Geiger mode and the device acronym is LmAPD.

## 2. LEONARDO LMAPDS

The physics of avalanche gain in HgCdTe has been well reported, for instance , and the references therein [1], [2], [3]. Also the HgCdTe Metal Organic Vapour Phase Epitaxy (MOVPE) technology employed by Leonardo for all its infrared detector products has been well described [5, 1]. The design and performance has evolved rapidly in recent years and it is worth summarizing the particular technology now used for Lm-APDs.

Figure 2 illustrates the history-dependent avalanche gain process in HgCdTe that provides the often quoted noise-free multiplication mechanism. However the noise figure relates only to gain fluctuations and does not include the noise due to dark current that might also be enhanced by the high bias voltage. The strategy for good APD design must include bandgap engineering (as illustrated in Figure 1) to suppress these currents.

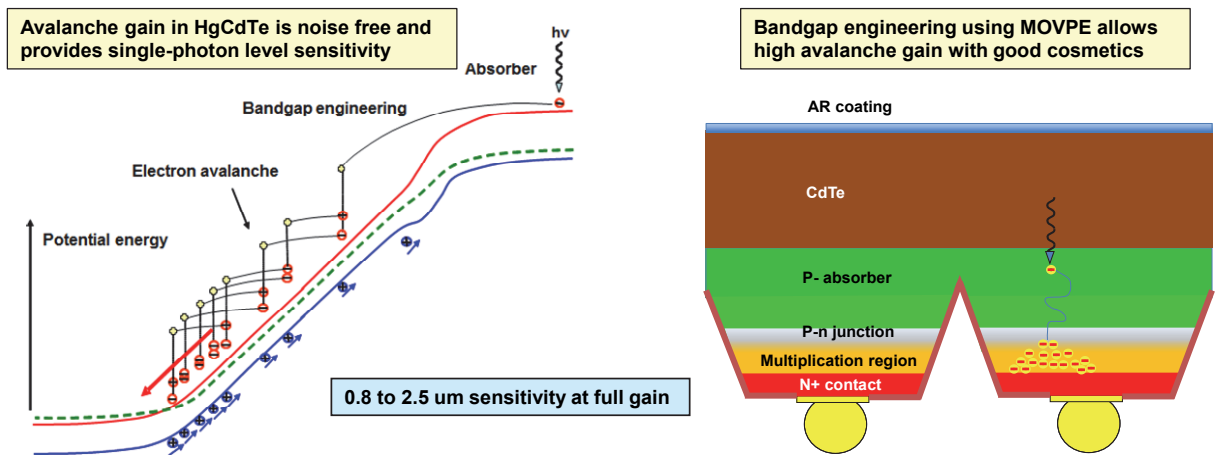

Figure 1:MOVPE design and device schematic

The main advantage of MOVPE for LmAPDs is the ability to incorporate complicated bandgap and doping profiles to switch off junction-related dark current sources, such as: trap-assisted tunnelling and generation-recombination current in the depletion layer. There is also a clear distinction between Liquid Phase Epitaxy (LPE) technologies grown on (111) CdZnTe and MOVPE which is grown close to (100) on GaAs. Homojunctions in LPE are susceptible to 111 threading dislocations and process induced dislocations intercepting the p-n junction. These appear to be absent or much reduced in MOVPE despite growth on a lattice miss-matched substrate. Instead defect revealing etches show a faint pattern of etch pits from weak mosaic crystal boundaries and these can be deactivated by widening the bandgap in critical areas. MOVPE allows for tight geometry control of individual layers so that the volume of narrow bandgap material in the structure is over an order less than planar or via-hole technologies and has a direct effect on thermal current. Bandgap engineering of the LmAPD can ensure that the dark current receives much lower avalanche gain that the photon signal, further suppressing the effect of dark current. In consequence the MOVPE mesa diode is intrinsically less sensitive to dark current mechanisms and pixel defects than other processes. Recent focus has been on bandgap profiles that mitigate tunnel-trap-tunnel currents that ultimately become the key limitation especially at low temperature.

# 3. THE C-RED ONE HERITAGE

C-RED1 is an ultra-low noise infrared camera based on the Saphira detector and fabricated by First Light Imaging, specialized in fast imaging camera, after the successful commercialization of the OCAM2 camera [4] dedicated to extreme adaptive optics wavefront sensing. Designed and fabricated by Leonardo UK, formerly Selex, the Saphira detector is designed for high speed infrared applications and is the result of a development program alongside the European Southern Observatory on sensors for astronomical instruments [5], [6], [7]. It delivers world leading photon sensitivity of <1 photon rms with Fowler sampling and high speed non-destructive readout (>10K frame/s). Saphira is an HgCdTe avalanche photodiode (APD) 320x256 array incorporating a full custom ROIC for applications in the 1 to 2.5µm range. C-RED One camera is an autonomous plug-and-play system with a user-friendly interface, which can be operated in extreme and remote locations. The sensor is placed in a sealed vacuum environ-ment and cooled down to cryogenic temperature using an integrated pulse tube. The vacuum is self-managed by the camera and no human intervention is required. Shown in the Figure 2, the camera has been extensively described in [8].

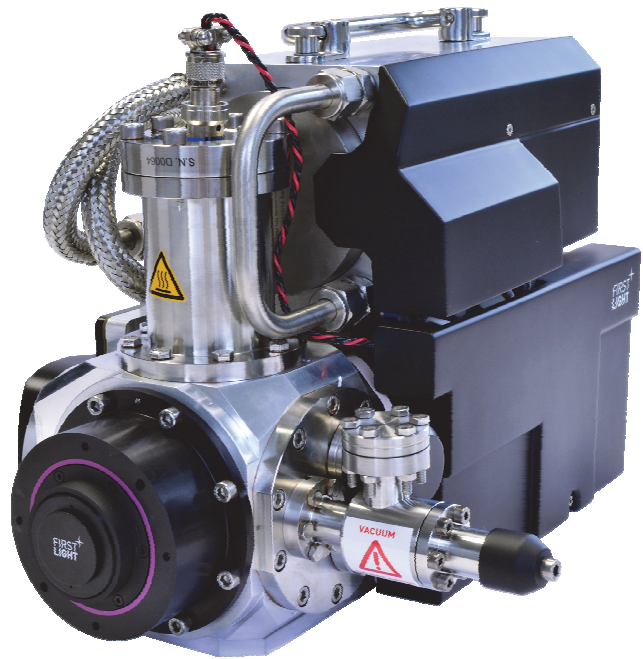

Figure 2: the C-RED One camera. The cooling system (pulse tube) can be seen on the top whereas in the bottom are the vacuum cryostat and the readout electronics.

The Table 1 shows the main C-RED One camera performances.

Table 1: C-RED one 320x256 SWIR APD camera performance

| TEST MEASUREMENT* | Result | Unit |
|---|---|---|
| Maximum speed Full Frame single readout | 3500 | FPS |
| Readout Noise at 1720 FPS CDS readout and gain x50, looking at a black body at a temperature of 90K | <1 | e- |
| Dark current looking at a black body at a temperature of 90K and e-APD gain x10 | <80 | e-/p/s |
| Quantization | 16 | Bit |
| Detector Operating Temperature (No LN2) | 90 | K |
| Flat Quantum Efficiency from 1.1 µm to 2.4 µm (J, H, K) at 100K | >60 | % |
| Operability due to signal response / pixels with signal <0.8*median at bias of 9V and integration time of 10 ms | <0.1 | % |
| Operability due to CDS noise / pixels with noise <2*median at bias of 9V and integration time of 10 ms | <0.1 | % |
| Excess noise Factor F | <1.25 | n/a |
| Pulse tube cooling, vibration imparted to the detector with respect to the front flange of the camera (RMS along each detector axis) | <1 | µm |

## 4. 2KX2K LMAPD DETECTOR DEVELOPMENT AT LEONARDO

### 4.1 ME1130 ROIC description

A new ROIC is currently developed by Leonardo to drive the future 2Kx2K LmAPD detectors under the name ME1130. The main characteristics of the ME1130 ROIC are given hereafter:

- 2048x2048 Format Array
- 4 Top + 4 Bottom Reference Rows
- 15m Pixel Pitch
- Low Glow Design
- Multiple windows. Both readout windows and reset windows are configurable
- 16, 8 and 4 Analog Video Outputs Selectable
- Output buffer or output source follower option
- Readout: Non-Destructive, Read-Reset-Read (row), interlace reference row, windowing
- Pixel by pixel and line by line reset
- Low intrinsic noise / Low Integration capacitance
- Low Voltage Operation (3.3V)

The schedule of the ROIC development is to receive and test the first engineering samples by September 2022 and hybridize the full specification array by end of 2022.

The Figure 3 shows the bloc diagram of the ME1130 ROIC. The Figure 4 shows a picture of the ROIC.

Figure 3: ME1130 2Kx2K LmAPD ROIC bloc diagram

Figure 4: ME1130 view

## 4.2 ME1130 ROIC specifications

The full ME1130 ROIC specifications are given in the Table 2.

Table 2: ME1130 ROIC full specifications

|        | **Parameter** | **Value** | **Units** | **Comments** | **Verification** |
|--------|---------------|-----------|-----------|--------------|------------------|
| REQ-1  | Number of Pixels | 2048 x 2048 | Pixels |  | Design |
| REQ-2  | Pixel size | 15 | µm | Square Pixel | Design |
| REQ-3  | Fil Factor | >90 | % | Active collection area | Design |
| REQ-4  | No of Outputs | 16 |  | Readout through 4,8 or 16 outputs shall be possible | Design/Test |
| REQ-5  | Read-Out mode | Global Shutter |  |  | Design/Test |
| REQ-6  | Frame Rate | ≥1 | Hz | Minimum frame rate for full frame readout through 16 outputs | Design/Test |
| REQ-7  | Integration time | Min: 10 Max: no limit | µsecs | Tint should be externally controlled | Design/Test |
| REQ-8  | Windowing |  |  | Multiple window readout – selectable window sizes, minimum is 1 window 1k x 1k | Design/Test |
| REQ -9 | APD gain | 1-1000 |  | Gain stability v bias voltage and operating temperature shall be characterised | Design/Characterisation/Test |
| REQ-10 | Cut On Wavelength | ≤0.8 | µm |  | Design |
| REQ-11 | Cut On Wavelength | ≥2.5 | µm |  | Design |
| REQ-12 | QE x Fill Factor | ≥70 | % | Assumes the appropriate AR coat, between 0.85 & 2.45 µm | Design / Characterisation |
| REQ-13 | Crosstalk | <4 | % | Summed total from all neighbouring pixels all wavelengths and up to full well fill | Characterisation |
| REQ-14 | Defects | 0.01% 0.1% 0 | Dead Pixels Defective pixels Column defects Row defects Clusters | For definition of a defect see below | Test |
| REQ-15 | Dark Current | <3 | e/p/s | At 80K & APD gain of 10 | Test |

| REQ-16 | DSNU | <10 | % | RMS | Test |
|---|---|---|---|---|---|
| REQ-17 | Read Noise | <10 | e- RMS | In dark conditions excluding dark current contribution. At an APD gain of 10, with CDS and full frame readout through 16 outputs | Test |
| REQ-18 | Excess Noise Factor | <1.2 | | At 80K, gain less than 30 | Test |
| REQ-19 | CHC | >100K | Ke- | Defined as the maximum number of charges storable within the linear regime of the detector (see Linearity criteria) | Test |
| REQ-20 | Non Linearity | <1 | % | The detectors response shall be a linear function of the input radiance. Any deviations from the linear function shall be less than 1% of the response corresponding to the local linear response in the interval 10-90% of full well capacity of the detector, for the whole spectral range. Non-linearity should be measured both verses incident flux and integration time | Test / Characterisation |
| REQ-21 | Operating Temperature | 80 | K | For testing at Leonardo | Test |
| REQ-22 | PRNU 6 x6 pixels Whole array | <0.33 | % | RMS | Test |
| REQ-23 | Voltage & power supplies | <3.3 | V | No external voltage higher than 3.3V shall be needed for the detector, with the exception of the diode bias | Design |
| REQ-24 | Power dissipation | <50 | mW | Full frame readout at 1Hz, 16 outputs | Design / Test |

## 4.3 Expected camera performances

The Table 3 shows the 2Kx2K LmAPD camera we expect to achieve.

Table 3: expected 2Kx2K LmAPD camera performances

| Parameter | Value |
|---|---|
| Readout Noise at 10 FPS and gain ~50, looking at a black body at a temperature of 60K | < 1 e- (sub e-) |
| Total background (dark current + thermal background) at 10 FPS and gain ~10 and 60K with J-band filter, looking at a black body at room temperature | ≤ 0.1-1 e-/s/pixel |
| Dark current at gain ~10 and 60K | < 0.01 e-/s/pixel |
| Quantization | 16bit |
| Detector Operating Temperature (No LN2) | <65K, 50-60K goal |
| Quantum Efficiency from 1.1 to 2.4 μm (J,H,K) | >60% |
| Excess noise Factor F | <1.25 |
| Pulse tube vibration imparted to the detector | <1 μm RMS |
| Filtering | Fixed long-wavelength suppression filters at front of the camera with cut-off wavelengths around 1.75 μm; Cold filter wheel, at least 6 positions (Y,J,H,cold stop, open position, custom optics) |

## 4.4 2Kx2K LmAPD camera development schedule

The engineering devices for the current program are expected by end of 2023. We expect to receive the science grade devices by end of 2024 and to build the first H band prototype camera by mid-2025. The camera with the filter wheel option is expected by end of 2025. The detector package and readout electronics for integration in any custom cryostat will also be offered, likely one year later.

## 5. CONCLUSION

LmAPD SWIR arrays are today moving from small formats fast readout to large format scientific imaging with long readout time. The First Light Imaging French company starts to develop an autonomous camera using the 2Kx2K LmAPD SWIR array currently developed by Leonardo and funded by ESA. This will be the first time a SWIR large format APD array format to be commercially available. This camera is a major technological breakthrough in the field of low noise scientific SWIR imaging. First Light Imaging intends to develop three camera variants: an autonomous camera with single cold filter and a custom beam aperture, the previous version with a cold filter wheel and detector package with its readout electronics and acquisition for integration inside any custom cryostat. The first cameras using this new device are expected by the second half of 2025.

## ACKNOWLEDGMENT

The large format 2Kx2K ROIC for LmAPD array is funded by ESA under a TDE program with the contract number 000130154/20/NL/AR.
## REFERENCES

[1] Kinch, M.A. and Baker, I.M, "Mercury Cadmium Telluride - Growth, Properties and Applications", editors: Capper, P. and Garland, J., Chapter 21 –"MCT Electron Avalanche Photodiodes", published by Wiley, UK., (2011).

[2] Rothman, J., Mollard, L., Bosson, S., *et al.,* "Short-wave infrared MCT avalanche photodiodes", *Journal of Electronic Materials*, **41** (10), 2928–2936., (2012).

[3] Beck, J.D., Scritchfield, R., Mitra, P., *et al*., "Advanced photon counting techniques"*,* SPIE, Conference Series, 8033, (2011).

[4] P. Feautrier et al., "OCam with CCD220, the Fastest and Most Sensitive Camera to Date for AO Wavefront Sensing", Publ. Astron. Soc. Pac. Vol 123 n°901, 263-274 (2011)

[5] J.L. Gach and P. Feautrier, "Electron initiated APDs improve high-speed SWIR imaging", Laser Focus World vol 51 n°9, 37-39, (2015)

[6] G. finger et al., "Evaluation and optimization of NIR HgCdTe avalanche photodiode arrays for adaptive optics and interferometry", Proc. SPIE 8453, 84530T (2012).

[7] Feautrier, Philippe, Gach, Jean-Luc, Wizinowich, Peter, "State of the art IR cameras for wavefront sensing using e-APD MCT arrays", AO4ELT4 Conference, 2015.  Rouvié, O. Huet, S. Hamard, JP. Truffer, M. Pozzi, J. Decobert, E. Costard, M. Zécri, P. Maillart, Y. Reibel, A. Pécheur "SWIR InGaAs focal plane arrays in France", SPIE Defense, Security and Sensing, 8704-2 (2013)

[8] Philippe Feautrier, Jean-Luc Gach, Timothée Greffe, Eric Stadler, Fabien Clop, Stephane Lemarchand, David Boutolleau, "C-RED One and C-RED 2: SWIR advanced cameras using Saphira e-APD and Snake InGaAs detectors", SPIE 10209, 102090G (2017)